\documentclass[aps,prd,twocolumn,nofootinbib]{revtex4-1}
\usepackage{epsfig}
\usepackage{amsmath}
\usepackage[colorlinks,linkcolor=blue,anchorcolor=blue,citecolor=blue,urlcolor=blue,breaklinks=true]{hyperref}
\usepackage{graphics}
\usepackage{color}
\usepackage{bibunits}

\begin{document}
\author{Peng Cheng$^{1}$, Qi Meng$^{1}$, Qian Wu$^{1}$, Jialun Ping$^{2,}$}~\email[]{Email: jlping@njnu.edu.cn}\author{Hongshi Zong$^{1,3,4,}$}~\email[]{Email: zonghs@nju.edu.cn}
\address{$^{1}$Department of Physics, Nanjing University, Nanjing 210093, China}
\address{$^{2}$Department of Physics, Nanjing Normal University, Nanjing 210097, China}
\address{$^{3}$Joint Center for Particle, Nuclear Physics and Cosmology, Nanjing 210093, China}
\address{$^{4}$State Key Laboratory of Theoretical Physics, Institute of Theoretical Physics, CAS, Beijing, 100190, China}

\title{Calculation of Dissociation Temperature of Nucleon Using Gaussian Expansion Method}
\begin{abstract}
The first study of the dissociation temperature of nucleon in hot QCD medium in the framework of constituent 
quark model is presented. The temperature-dependent potential energy of the three quark system, taking as 
the internal energy of the system are obtained from the free energy of the system, and the temperature-dependent 
free energy is derived based on Debye-H\"uckel theory. The lattice QCD results of the free energy for heavy 
three-quark system are employed and extended to the light three-quark system. The Schr\"{o}dinger equation for 
nucleon is solved with the help of Gaussian expansion method and the dissociation temperature of the nucleon
is determined according to the temperature dependence of binding energy and radii. Comparing with the dissociation 
temperature of $J/\psi$, the dissociation temperature of nucleon is higher. So, nucleon is more difficult to melt 
than Charmonium.

\noindent $Keywords$: Nucleon; Dissociation temperature; Gaussian Expansion Method.
\end{abstract}

\pacs{12.38.Mh, 12.39.-x, 25.75.Nq}

\maketitle

\section{Introduction}
The relativistic heavy-ion collider experiments show that human beings may have produced quark-gluon plasma (QGP) in the laboratory \cite{torres2015flavor}.
It is generally believed that some quark bound states can survive in QGP.
In thermal Quantum chromodynamics (QCD), the thermal 
properties of QGP can be determined by studying the behavior of these quark bound states in hot medium. 
In 1986, Satz pointed out that the suppression of $J/\psi$ could be a signature of QGP formation 
in the relativistic heavy ion collisions \cite{matsui1986j}. Since then, many people have systematically studied
the melting of Charmonium and Bottomonium. 
But the dissociation temperature of nucleon, the lightest baryon, has not been studied systematically.
This is due to the difficulties of solving three-body system and obtaining the temperature dependent quark-quark interaction potential within the nucleon.

QCD is the fundamental theory of strong interaction. It works well in the 
perturbative region but it does not work in the non-perturbative region. It is difficult 
for us to use QCD to study the thermal properties of quark bound states directly. 
In this case, people have to use the model \cite{karsch1988color,satz2006colour,alberico2005heavy,zhen2012dissociation} to study the dissociation of 
quarkonium states. In high temperatures and density, the interaction between quarks 
is screened \cite{digal2005heavy} and the binding energy will be decreased. As a result, the nucleon 
will start to melt when the binding energy become low enough. The melting of nucleon can be solved by Schr\"odinger equation of three body. 
For the calculation, we need the interacting potentials among 
quarks of the nucleon in the hot medium, which is temperature-dependent. Unfortunately, the potentials 
are not yet well understood up to now. The free energy of a static heavy three-quark system $F_{qqq}(r,T)$ can be 
calculated in lattice QCD and the internal energy can be obtained by using thermodynamic relation. 
In the present approach, the needed potential is assumed to be the internal energy, i.e. $V=F+sT$ with 
$s$ being the entropy density $s=-\partial F/\partial T$. The temperature-dependent form of 
$F_{q\bar q}(r,T)$ can be constructed based on Debye-H\"uckel theory \cite{dixit1990charge}, 
as having been done in Ref. \cite{digal2005heavy}. Then, we can determine the T-dependent parameter 
in the free energy by fitting it with the lattice data. According to the relation between 
$F_{q\bar q}(r,T)$ and $F_{qqq}(r,T)$, we can obtain the free energy of heavy three-quark system. 
We assume the conclusions are applicable to light quark system due to the flavor independence of the strong interaction. After constructing the potential 
of nucleon system at finite temperature, we can obtain the temperature dependence of binding energies 
and radii by solving the corresponding Schr\"odinger equation. The dissociation temperature is 
the point where the binding energy decreases to zero. The Gaussian expansion method (GEM), which
is an efficient and powerful method in few-body system \cite{hiyama2003gaussian}, is employed
to calculate dissociation temperature of nucleon in this paper. 

This paper is organized as follows. In Sec.\ref{sec:quakonium}, we show the rationality of GEM 
on studying the melting of Charmonium and Bottomonium
by comparing our results  with others. In Sec.\ref{sec:forma}, we construct the potential of nucleon 
and apply GEM to solve corresponding Schr\"odinger equation. In Sec.\ref{sec:res}, we show the 
results at quenched and 2-flavor QCD, respectively. Sec.\ref{sec:summ} contains summary and conclusion.

\section{The Results on Dissociation Temperature of Quarkonium}\label{sec:quakonium}

 Before studying the dissociation of nucleon, we test the rationality of GEM on studying dissociation temperatures of quarkonium
 by comparing our results with others. To compare with Satz's results, the potential of quarkonium at 
 finite temperature we use is the same as Satz's work \cite{satz2006colour}. The results on dissociation 
 temperature of Charmonium, Bottomonium in Ref.\cite{satz2006colour} and our calculation results are listed 
 in Table\ref{tab:temp}, Table\ref{tab:temp2},
\begin{table}[!h]
\begin{center}
\caption{\label{tab:temp}Dissociation Temperature $T_d/T_c$ of Charmonium in Ref.\cite{satz2006colour} and our results.}
\renewcommand\arraystretch{1.8}
\begin{tabular}{cccc}
\hline
Charmonium   &  1S  &  1P  &  2S \\
\hline
Ref.\cite{satz2006colour} & 2.10 & 1.16 & 1.12 \\
\hline
Our Results & 2.08 & 1.16 & 1.14   \\
\hline
\end{tabular}
\end{center}
\end{table}
\begin{table}[!h]
\begin{center}
\caption{\label{tab:temp2}Dissociation Temperature $T_d/T_c$ of Bottomonium in Ref.\cite{satz2006colour} and our results.}
\renewcommand\arraystretch{1.8}
\begin{tabular}{cccccc}
\hline
Bottomonium   &  1S  &  1P  &  2S & 2P & 3S \\
\hline
Ref.\cite{satz2006colour} & $>$ 4.0 & 1.76 & 1.60 & 1.19 & 1.17 \\
\hline
Our Results & 5.83 & 1.72 & 1.59 & 1.18 & 1.17\\
\hline
\end{tabular}
\end{center}
\end{table}
which show our results are consistent with that in Ref.\cite{satz2006colour}. So GEM can give accurate results 
on the dissociation temperature of quarkonium. Giving accurate binding energy and wave function \cite{hiyama2003gaussian} 
makes GEM very suitable for studying dissociation temperature of quark bound states (more detail can be found in Appendix A). In the following, we will use this method to 
calculate the dissociation temperature of nucleon.

\section{Formalism}\label{sec:forma}

\subsection{Constituent Quark Model}

The constituent quark model is a non-relativistic quark model \cite{yang2008dynamical}. In the constituent quark model, 
baryons are formed by three constituent quarks, which are confined by a confining potential and interact with each 
other \cite{yoshida2015spectrum}. The potential of baryon can be described by a sum of the potential of corresponding 
two-quark system. In Kaczmarek's work \cite{PhysRevD.75.054504}, it has been calculated in lattice QCD that the 
potential of diquark system is about half of that of corresponding quark-antiquark system, i.e. 
$V_{q q}=\frac 12 V_{q \bar q}$. The simplest and most frequently used potential for a $q \bar q$ system is 
the Cornell potential \cite{digal2005heavy},
\begin{equation}
    V_{q \bar q}(r) = -\frac{\alpha}{r} + \sigma r
\label{eq:cornel}
\end{equation}
where $\alpha$ is the coupling constant, and $\sigma$ is the string tension. In the present work, we neglect the 
spin-dependent part of potential here. Thus our Hamiltonian is written as
\begin{equation}
H = \sum_{i=1}^3 (m_i + \frac{\boldsymbol{p}_i^2}{2m_i}) - T_{cm} + \sum_{1=i<j}^3 \frac 12 V(r_{ij})
\label{eq:hamil}
\end{equation}
\begin{equation}
V(r_{ij}) = \sigma r_{ij} - \frac{\alpha}{r_{ij}}
\label{eq:cor1}
\end{equation}
where $m_i$ is the constituent quark mass of the $i$-th quark, and $T_{cm}$ is the kinetic energy of center-of-mass frame (cm). $\boldsymbol{r}_{ij}=\boldsymbol{r}_i-\boldsymbol{r}_j$ is the relative coordinate between $i$-th quark and 
$j$-th quark. In this model, the mass of light quark (u and d quark) we use is $300$ MeV. The parameters of 
Cornell potential we use are: $\alpha=1.4$, $\sqrt{\sigma}=0.131$ GeV. Solving the corresponding Schr\"odinger 
equation, $ H\Psi_{total}=E_m \Psi_{total}$, with GEM, we can get the mass $E_m$ and corresponding wave function 
of nucleon $\Psi_{total}$. We define the radii of nucleon as
\begin{equation}
R=\frac 13 \sum_{i=1}^3 \sqrt{\langle r_i^2 \rangle}
\label{eq:radii}
\end{equation}
with
\begin{equation}
\langle r_i^2 \rangle=\int \Psi^*_{total} r_i^2 \Psi_{total} d\tau
\label{eq:radii1}
\end{equation}
where $r_i$ is the distance between the center of nucleon and $i$-th quark. Using the calculated wave function, 
we can calculate the radii of nucleon. The calculating mass and radii of nucleon are $939$ MeV and $0.83933$ fm, 
respectively. While the corresponding experimental data are about $939$ MeV and $0.841$ fm. We can see this model 
gives a good estimation of the properties of nucleon even if the spin-dependent part is neglected. 
So it is reasonable for us to use this potential model to study the dissociation of nucleon. Of course, 
we need notice that the spin-dependent part plays an important role in the baryon spectrum.

\subsection{Wave Function}

\begin{figure}[!htbp]
\centering
\includegraphics[width=0.4\textwidth]{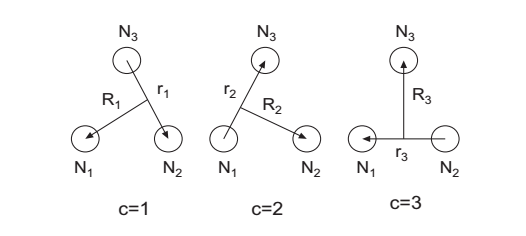}
\caption{\label{fig:jacobi} Three sets of Jacobi coordinates for a three-body system \cite{hiyama2003gaussian}.}
\end{figure}

Here, we solve the Schr\"odinger equation with GEM. In this method, three sets of Jacobi 
coordinates (Fig.\ref{fig:jacobi}) are introduced to express three-quark wave function. The Jacobi coordinates in
each channel $c (c=1,2,3)$ are defined as
\begin{equation}
\boldsymbol{r}_i=\boldsymbol{x}_j-\boldsymbol{x}_k
\label{eq:recoor1}
\end{equation}
\begin{equation}
\boldsymbol{R}_i=\boldsymbol{x}_i-\frac {m_j\boldsymbol{x}_j+m_k\boldsymbol{x}_k}{m_j+m_k}
\label{eq:recoor2}
\end{equation}
where $\boldsymbol{x}_i$ is the coordinate of the $i$-th quark and $(i,j,k)$ are given by Table~\ref{tab:jacobi}.
\begin{table}[!h]
\begin{center}
\caption{\label{tab:jacobi}The quark assignments $(i,j,k)$ for Jacobi channels.}
\renewcommand\arraystretch{1.8}
\begin{tabular}{cccc}
\hline
channel   &  i  &  j  &  k \\
\hline
1 & 1 & 2 & 3 \\
\hline
2 & 2 & 3 & 1 \\
\hline
3 & 3 & 1 & 2 \\
\hline
\end{tabular}
\end{center}
\end{table}
The total wave function is given as a sum of three rearrangement channels ($c=1-3$)
\begin{equation}
\Psi_{total}^{JM} = \sum_{c,\alpha} C_{c,\alpha} \Psi_{JM,\alpha}^{(c)}(\boldsymbol{r}_c,\boldsymbol{R}_c)
\label{eq:totalwave}
\end{equation}
where the index $\alpha$ represents ($s,S,l,L,I,n,N$). Here $s$ is the spin of the $(i,j)$ quark pair, 
$S$ is the total spin, $l$ and $L$ are the orbital angular momentum for the coordinate $\boldsymbol{r}$ 
and $\boldsymbol{R}$, respectively, and I is the total orbital angular momentum. The wave function for 
channel $c$ is given by
\begin{equation}
\Psi_{JM,\alpha}^{(c)}(\boldsymbol{r}_c,\boldsymbol{R}_c) = \phi_c \otimes [X_{S,s}^{(c)} 
\otimes \Phi_{l,L,I}^{(c)}]_{JM} \otimes H_{T,t}^{(c)}
\end{equation}
as given in Ref. \cite{yoshida2015spectrum}. The orbital wave function $\Phi_{l,L,I}^{(c)}$ is given 
in terms of the Gaussian basis functions written in Jacobi coordinates $\boldsymbol{r}_c$ and $\boldsymbol{R}_c$
\begin{equation}
\Phi_{l,L,I}^{(c)} = [\phi_l^{(c)}(\boldsymbol{r}_c) \psi_L^{(c)}(\boldsymbol{R}_c)]_I
\label{eq:orbital}
\end{equation}
\begin{equation}
\phi_{lm}^{(c)}(\boldsymbol{r}_c) = N_{nl} r_c^l e^{-\nu_n r_c^2} Y_{lm}(\hat{\boldsymbol{r}}_c)
\label{eq:orbital1}
\end{equation}
\begin{equation}
\psi_{LM}^{(c)}(\boldsymbol{R}_c) = N_{NL} R_c^L e^{-\lambda_N R_c^2} Y_{LM}(\hat{\boldsymbol{R}}_c)
\label{eq:orbital2}
\end{equation}
where the range parameters, $\nu_n$ and $\lambda_N$, are given by
\begin{gather}
\nu_n = 1/{r_n^2}, r_n = r_1 a^{n-1} (n=1,...,n_{max}), \notag \\
\lambda_N = 1/{R_N^2}, R_N = R_1 A^{N-1} (N=1,...,N_{max});
\label{eq:rangepara}
\end{gather}
In Eqs. (\ref{eq:orbital1}) and (\ref{eq:orbital2}), $N_{nl}$($N_{NL}$) \cite{hiyama2003gaussian} denotes 
the normalization constant of Gaussian basis. The coefficients $C_{c,\alpha}$ of the variational 
wave function, Eq. (\ref{eq:totalwave}), are determined by Rayleight-Ritz variational principle.

\subsection{Potential model for nucleon}
\begin{figure}[!htbp]
\centering
\includegraphics[width=0.4\textwidth]{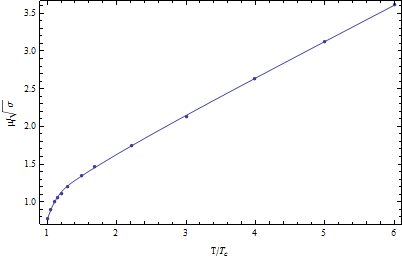}
\includegraphics[width=0.4\textwidth]{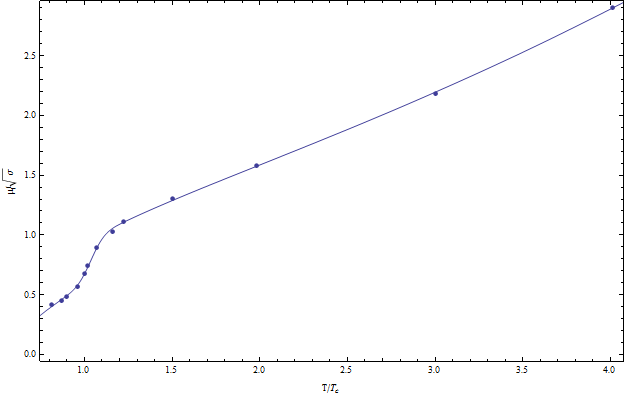}
\caption{\label{fig:miu} Results for $\mu(T)$ in quenched (upper figure) and 2-flavor (lower figure) QCD.}
\end{figure}

\begin{figure}[!htbp]
\centering
\includegraphics[width=0.4\textwidth]{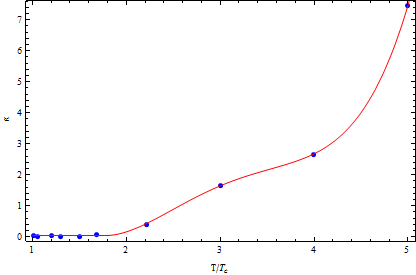}
\includegraphics[width=0.4\textwidth]{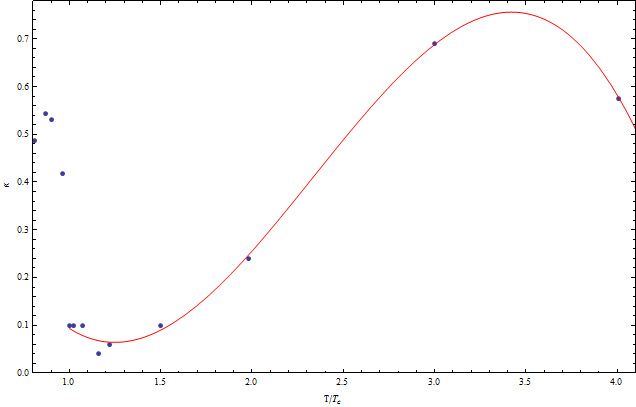}
\caption{\label{fig:kai}Results for $\kappa(T)$ in quenched (upper figure) and 2-flavor (lower figure) QCD.}
\end{figure}

\begin{figure}[!htbp]
\centering
\includegraphics[width=0.4\textwidth]{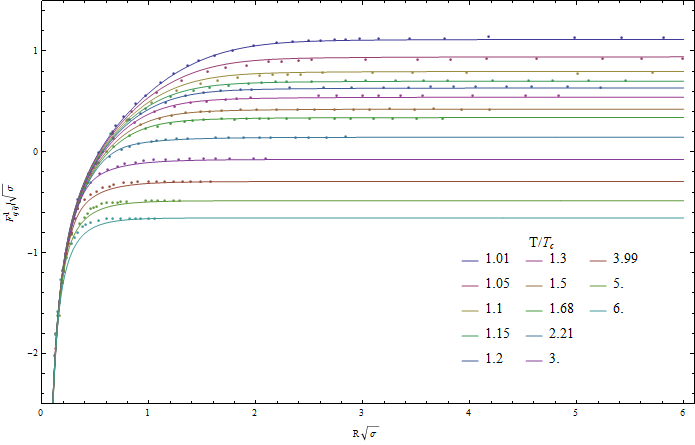}
\includegraphics[width=0.4\textwidth]{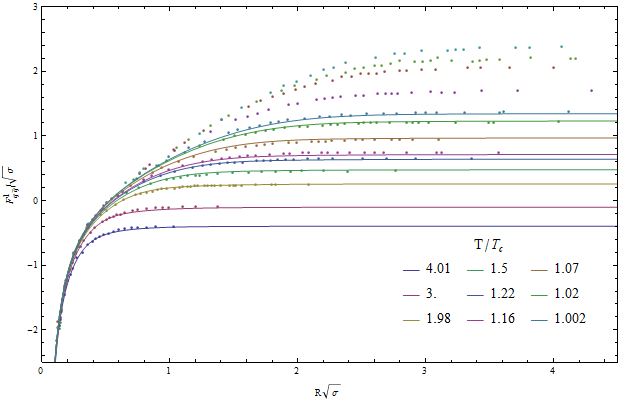}
\caption{\label{fig:lattfree}Results for free energy ($F_{q\bar{q}}^1$) in quenched (upper figure) and 2-flavor (lower figure) QCD.}
\end{figure}

The potential of nucleon at zero temperature has been discussed above and its parameters have been determined by 
fitting the properties of nucleon. To determine the dissociation temperature of nucleon, we need the potential in 
hot medium, i.e. $V_{qqq}(\boldsymbol{r},T)$ (the index q represents u or d quark). Here, we assume that 
the potential is just the internal energy
\begin{equation}
\begin{split}
V_{qqq}(\boldsymbol{r},T) &= U_{qqq}(\boldsymbol{r},T) \\
                          &= F_{qqq}(\boldsymbol{r},T) + sT
\end{split}
\label{eq:poten}
\end{equation}
where $s$ is the entropy density $s=-\partial F_{qqq}/\partial T$. 
In Refs.~\cite{hubner2008heavy,hubner2005free,huebner2005heavy}, Kaczmarek's works show that the color singlet 
free energies of the heavy three-quark system ($F_{qqq}^1$) can be described by the sum of antitriplet 
free energies of the corresponding diquark system ($F_{qq}^{\overline 3}$) plus self energy contributions 
when the temperature is above $T_c$. It can be expressed as
\begin{equation}
F_{qqq}^1(P,T) \simeq \sum_{i<j} F_{qq}^{\overline 3}(R_{ij},T)-3F_q(T)
\label{eq:free}
\end{equation}
where $P=\sum_{i<j}R_{ij}$ and the self energy $F_q(T)=\frac 12 F_{qq}^{\overline 3}(\infty,T)$. In Ref.~\cite{PhysRevD.75.054504}, O.Kaczmarek's work suggests a simple relation between free energies of anti-triplet $qq$ states and color singlet $q\bar{q}$
\begin{equation}
F_{q\bar{q}}^1(r,T) \simeq 2(F_{qq}^{\overline 3}(r,T)-F_q(T))
\label{eq:free1}
\end{equation}
The form of $F_{q\bar{q}}^1$ can be obtained based on studies of screening in Debye-H\"uckel theory. 
It can be written as \cite{digal2005heavy}
\begin{eqnarray}
F_{q\bar{q}}^1(r,T)=-\frac{\alpha}{r}\left[e^{-\mu r}+\mu r\right]+\frac{\sigma}{\mu}[\frac{\Gamma\left(1/4\right)}{2^{3/2}
    \Gamma\left(3/4\right)}            \notag \\
    -\frac{\sqrt{\mu r}}{2^{3/4}\Gamma\left(3/4\right)}K_{1/4}\left[\left(\mu r\right)^2+\kappa \left(\mu r\right)^4\right]]
    \label{eq:free2}
\end{eqnarray}
where screening mass $\mu$ and the parameter $\kappa$ are temperature-dependent, and $K_{1/4}[x]$ is the modified 
Bessel function. We can determine the T-dependent $\mu$ and $\kappa$ by fitting $F_{q\bar{q}}^1(r,T)$ to the 
lattice result obtained in quenched \cite{kaczmarek2002heavy} and 2-flavor \cite{kaczmarek2005static} QCD. 
At $r=\infty$, the free energy $F_{q\bar{q}}^1(T)$ is wirtten as
\begin{equation}
F_{q\bar{q}}^1(T) = \frac {\sigma}{\mu\left(T\right)} \frac {\Gamma\left(1/4\right)}{2^{3/2} \Gamma\left(3/4\right)} - \alpha \mu\left(T\right)
\label{eq:inffree}
\end{equation}
Thus, the form of $\mu(T)$ is given as function of $F(T)$
\begin{equation}
\mu\left(T\right) = \frac {\left[\sqrt{{F_{q\bar{q}}^1(T)}^2+4\sigma\alpha \frac {\Gamma\left(1/4\right)}{2^{3/2} \Gamma\left(3/4\right)}}-F_{q\bar{q}}^1(T)\right]}{2\alpha}
\label{eq:fmu}
\end{equation}
Once we obtain the temperature dependence of $\mu(T)$, we fit Eq.~(\ref{eq:free2}) to the lattice data to obtain 
$\kappa(T)$. The results for $\mu(T)$ and $\kappa(T)$ are shown in Fig. \ref{fig:miu} and Fig. \ref{fig:kai}, 
respectively. In Fig.~\ref{fig:lattfree}, we show our fit curves (solid lines) together with the lattice results. 
We can see that the resulting $F_{q\bar q}^1(r,T)$ fits the lattice results quite well for all $r$ and in a broad 
range of temperatures from $T_c$ to $4T_c$ in the two cases. For higher temperature, the resulting $F_{q\bar q}^1(r,T)$ 
cannot be fitted quite well to the lattice results in quenched QCD. There can be higher order corrections to 
Poisson equation \cite{digal2005heavy}.

To obtain the binding energies of nucleon, we define a effective potential as
\begin{equation}
\tilde V_{qqq}(\boldsymbol{R} ,T) = V_{qqq}(\boldsymbol{R} ,T) - V_{qqq}(\infty,T)
\label{eq:effpoten}
\end{equation}
Combining Eqs.(\ref{eq:poten}-\ref{eq:free1},\ref{eq:effpoten}), we can get a relation between effective potential 
and free energies of $q\bar q$
\begin{equation}
\tilde V_{qqq}(\boldsymbol{R},T) = \sum_{i<j} \frac 12 (\tilde F_{q\bar{q}}^1(R_{ij},T) - T\frac {\partial \tilde F_{q\bar{q}}^1(R_{ij},T)}{\partial T})
\label{eq:effpoten1}
\end{equation}
where
\begin{equation}
\tilde F_{q\bar{q}}^1(R_{ij},T) = F_{q\bar{q}}^1(R_{ij},T) - F_{q\bar{q}}^1(\infty,T)
\label{eq:efffree}
\end{equation}
Replacing the potential term, $\sum_{1=i<j}^3 \frac 12 V(r_{ij})$, in Eq.~(\ref{eq:hamil}) with this effective potential 
$\tilde V_{qqq}(\boldsymbol{R},T)$, we can get a new Hamiltonian for nucleon at finite temperature written as
\begin{equation}
H_{new} = \sum_{i=1}^3 \frac{\boldsymbol{p}_i^2}{2m_i} - T_{CM} + \tilde V_{qqq}(\boldsymbol{R},T)
\label{eq:hamil1}
\end{equation}
Solving corresponding Schr\"odinger euqation, 
$$ 
H_{new}\Psi_{total}^{JM}=\epsilon(T) \Psi_{total}^{JM},
$$ 
with GEM, we can get the binding energies $\Delta E(T)(=-\epsilon(T))$ and corresponding wave function at finite temperature. 
Using the wave function, we can calculate the T-dependent radii according to Eq.~(\ref{eq:radii}).

\section{Numerical Results}\label{sec:res}
\begin{figure}[!htbp]
\centering
\includegraphics[width=0.4\textwidth]{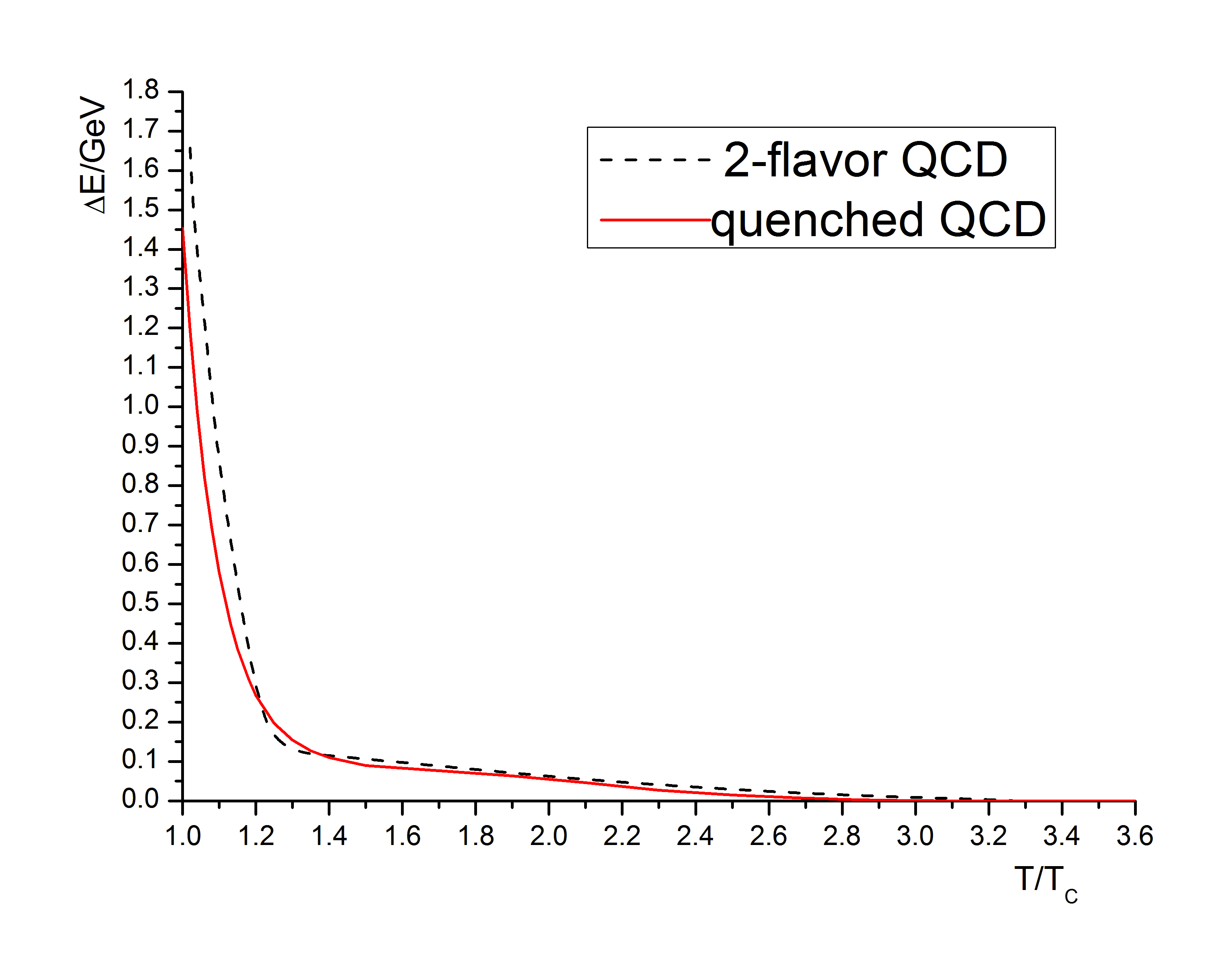}
\caption{\label{fig:bind}T-dependent of binding energy in quenched and 2-flavor QCD, respectively.}
\end{figure}

\begin{figure}[!htbp]
\centering
\includegraphics[width=0.4\textwidth]{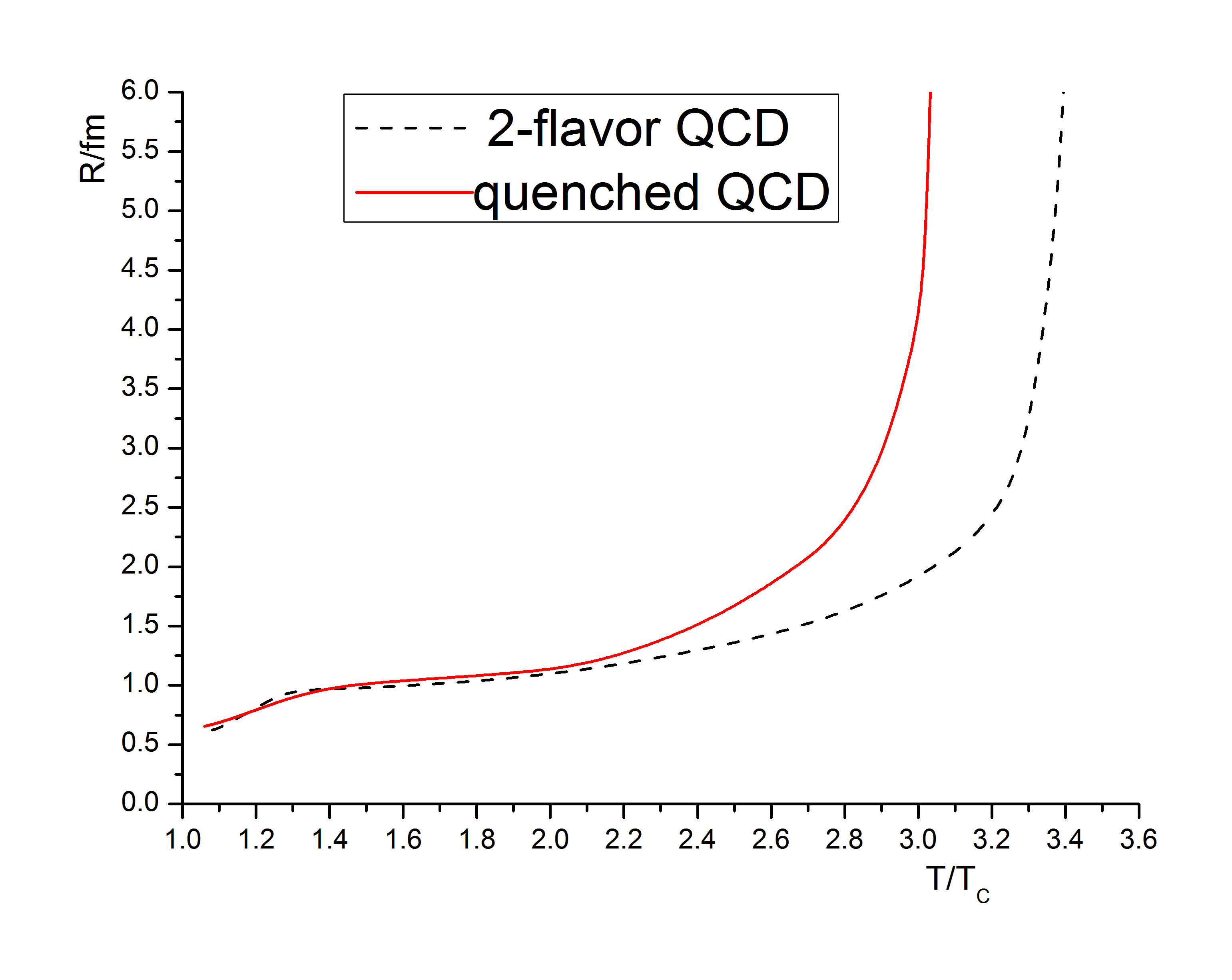}
\caption{\label{fig:ra}T-dependent of radii. in quenched and 2-flavor QCD, respectively}
\end{figure}

In Fig. \ref{fig:bind}, we show the resulting binding energies behaviour for nucleon in quenched and 2-flavor QCD. 
We can see there is little difference between the two lines. When they vanish, the nucleon no longer exists. 
So $\Delta E(T)=0$  determines the dissociation temperature. From Fig. \ref{fig:bind}, we get the dissociation 
temperature in quenched and 2-flavor QCD are about $3.0T_c$ and $3.3T_c$, respectively; in Fig. \ref{fig:ra}, 
we show the corresponding nucleonic radii. The dissociation temperature determined from Fig. \ref{fig:ra} is consistent 
with that determined from Fig. \ref{fig:bind}. It is seen that the divergence of the radii defines quite well 
the different dissociation points in the two cases. The resulting dissociation temperatures have a little 
difference between the two cases.

\section{Summary and Conclusion}\label{sec:summ}
The free energies of quark-antiquark system we construct based on Debye-H\"uckel theory at finite temperature 
fit the lattice results quite well from $T_c$ to $4T_c$, but not well for higher temperature. According to 
Kaczmarek's works, we can get a relation between color singlet free energy of heavy $qqq$ system $F_{qqq}^1$ 
and color singlet free energy of heavy $q\bar q$ system $F_{q\bar q}^1$, written as $F_{qqq}^1 \simeq \sum_{i<j} 
\frac 12 F_{q\bar q}^1$. The dissociation temperature of nucleon in quenched and 2-flavor QCD we calculate are 
about $3.0T_c$ and $3.3T_c$, respectively. There are a little difference between the two results. Comparing with 
$J/\psi$, the dissociation temperature of nucleon is higher. So, nucleon is more difficult to melt than charmonium. 
For the potential, we neglect the spin-dependent part in this work which may has some effects to the resulting 
dissociation temperature. The effects arising from spin-dependent part deserve further studies.

\section*{Acknowledgement}

We are grateful to Emiko Hiyama, Pengfei Zhuang, Yunpeng liu, Min He and Fan Wang for their work and helpful discussions.

This work is supported in part by the National Natural Science Foundation of China (under Grants Nos. 11475085, 11535005, 
11690030 and 11775118), the Fundamental Research Funds for the Central Universities (under Grant No. 020414380074), 
and the International Science \& Technology Cooperation Program of China (under Grant No. 2016YFE0129300).

\bibliography{ref}
\begin{appendix}
\section{The Calculation on Dissociation Temperature of Quarkonium}
\begin{figure}[!htbp]
\centering
\includegraphics[width=0.4\textwidth]{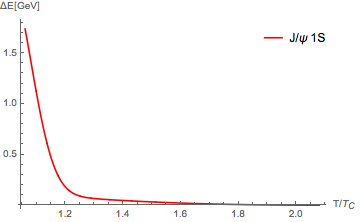}
\includegraphics[width=0.4\textwidth]{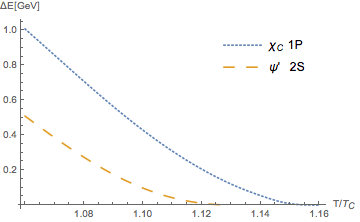}
\caption{\label{fig:charm}Binging energy of $J/\psi (1S)$, $\chi_c(1P)$, and $\psi'(2S)$ dependent to $T$}
\end{figure}

\begin{figure}[!htbp]
\centering
\includegraphics[width=0.4\textwidth]{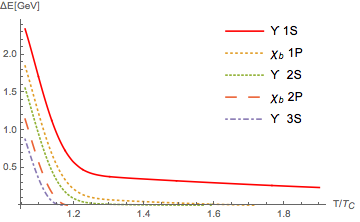}
\caption{\label{fig:bottom}Binging energy of $\Upsilon (1S)$, $\chi_b(1P)$, $\Upsilon (2S)$, $\chi_b(2P)$, and $\Upsilon (3S)$ dependent to $T$}
\end{figure}

\begin{figure}[!htbp]
\centering
\includegraphics[width=0.4\textwidth]{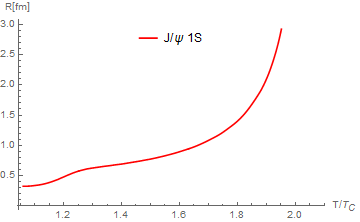}
\includegraphics[width=0.4\textwidth]{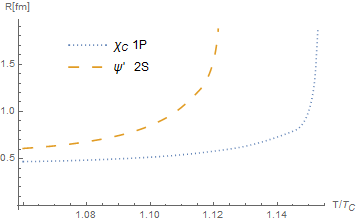}
\caption{\label{fig:RC} Bound state radii of $J/\psi (1S)$, $\chi_c(1P)$, and $\psi'(2S)$ dependent to $T$}
\end{figure}

\begin{figure}[!htbp]
\centering
\includegraphics[width=0.4\textwidth]{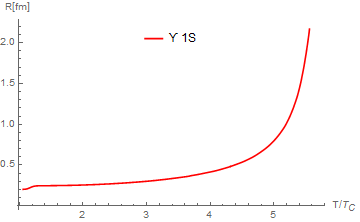}
\includegraphics[width=0.4\textwidth]{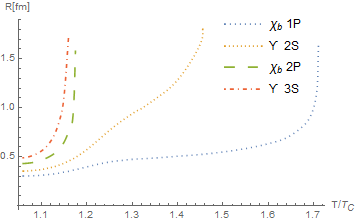}
\caption{\label{fig:RB} Bound state radii of $\Upsilon (1S)$, $\chi_b(1P)$, $\Upsilon (2S)$, $\chi_b(2P)$, and $\Upsilon (3S)$ dependent to $T$}
\end{figure}

We have given our results on dissociation temperature of quarkonium in Sec.\ref{sec:quakonium}. Here we present the calculation in detail. For quarkonium, the potential we use is the Cornell potential mentioned above. The Hamiltonian for quarkonium is written as
\begin{equation}
H = \frac{\boldsymbol{p}^2}{2\mu_{12}} + V_{q\bar q}(\boldsymbol{r})
\label{eq:hamila}
\end{equation}
where $\boldsymbol{p}=\boldsymbol{p}_2-\boldsymbol{p}_1$ and $\mu_{12}$ is the reduced mass. The parameters taken from Ref.\cite{satz2006colour} are: $m_c=1.25$ GeV, $\sqrt{\sigma}=0.445$ GeV, $\alpha=\pi/12$ and $m_b=4.65$ GeV. Then we can construct the corresponding free energies at finite temperature based on Debye-H\"uckel theory as having done above. To compare with H.Satz's work, we neglect the term $\kappa \left(\mu r\right)^4$ in Eq. (\ref{eq:free2}). We can obtain the T-dependent parameter $\mu(T)$ by fitting Eq. (\ref{eq:free2}) to lattice data calculated in two-flavour QCD\cite{kaczmarek2005static}. According to the potential model mentioned above, we can construct the potential of quarkonium at finite temperature. After obtaining the potential of quarkonium, we can get the Schr\"{o}dinger equation
 \begin{eqnarray}
    \left(-\frac{1}{2\mu_{12}}\nabla^2+V_{q\bar q}(r,T)-V_{q\bar q}(\infty,T)\right)\psi_i (r,T) \notag\\
    =\epsilon_i (T)\psi_i (r,T)
    \end{eqnarray}
where the index $i$ represents a quarkonium state and $\Delta E_i(T)(=-\epsilon_i (T))$ is the binding energy at temperature T. According to Ref.~\cite{hiyama2003gaussian}, we expand the total wave function in terms of a set of basis functions as
\begin{equation}
    \psi_{lm}=\sum_{n=1}^{n_{max}}c_{n}\phi_{nlm}
    \end{equation}
with
\begin{equation}
    \phi_{nlm}(r)=\phi_{nl}(r)Y_{lm}(\hat{\boldsymbol{r}})
 \end{equation}
 \begin{equation}
    \phi_{nl}(r)=N_{nl}r^l e^{-v_n r^2}
 \end{equation}
where the $N_{nl}$ is the normalization constant. And Rayleight-Ritz variational principle leads to a generalized matrix eigenvalue problem,
    \begin{equation}
    \sum_{n'=1}^{n_{max}}\left(H_{nn'}-EN_{nn'}\right)c_{n'l}=0
    \end{equation}
Therefore, we can obtain the eigenvalues and corresponding wave functions of both ground state and excited states. We define the radii for quarkonium as:
\begin{eqnarray}
     \langle r_i \rangle=\int \psi^*_i r \psi_i d\tau
\end{eqnarray}
Then we can use the calculating wave function to calculate the temperature-dependent radii.
In Fig.~\ref{fig:charm} and Fig.~\ref{fig:bottom}, we show the resulting binding energy behaviour for the different charmonium states and bottomonium states, respectively. We show the T-dependence of bound state radii for each states in Fig.~\ref{fig:RC} and Fig.~\ref{fig:RB}. According to these figures, we can determine the dissociation temperature.


\end{appendix}
\end{document}